\documentclass[12pt,preprint]{aastex}
\usepackage{psfig}
\usepackage{emulateapj5,apjfonts,times}

\newcommand{\beq}{\begin{equation}}
\newcommand{\eeq}{\end{equation}}

\def\alp{\mbox{$\alpha$}}

\def\earth{\hbox{$\oplus$}}

\def\solar{\mbox{$_{\normalsize\odot}$}}

\newcommand{\AmS}{{\protect\the\textfont2
  A\kern-.1667em\lower.5ex\hbox{M}\kern-.125emS}}
\newcommand{\lsim}{\ \raise
-2.truept\hbox{\rlap{\hbox{$\sim$}}\raise5.truept\hbox{$<$}\ }}
\newcommand{\gsim}{\ \raise
-2.truept\hbox{\rlap{\hbox{$\sim$}}\raise5.truept\hbox{$>$}\ }}
\newcommand{\simsim}{\ \raise
-2.truept\hbox{\rlap{\hbox{$\sim$}}\raise5.truept\hbox{$\sim$}\ }}

\slugcomment{Accepted for publication in ApJ Letters}

\righthead{}
\lefthead{D. Gouliermis,\ W. Brandner \& Th. Henning}

\shorttitle{Pre-Main Sequence population of the association LH 52 in the LMC}
\shortauthors{Gouliermis D., Brandner W. \& Henning Th.}

\begin{document}

\title{The low-mass pre-main sequence population of the stellar 
association LH 52 in the Large Magellanic Cloud discovered with Hubble 
Space Telescope WFPC2 Observations}


\author{D. Gouliermis, W. Brandner, Th. Henning}
\affil{Max-Planck-Institut f\"{u}r Astronomie, K\"{o}nigstuhl
17, D-69117 Heidelberg, Germany}
\email{dgoulier@mpia.de, brandner@mpia.de, henning@mpia.de}


\begin{abstract} 
We report on the serendipitous discovery of $\sim$ 500 low-mass candidate
PMS stars in the vicinity of the stellar association LH 52 in the Large
Magellanic Cloud. We present evidence that the red faint sequence of these
stars seen in the CMD of LH 52 from HST/WFPC2 observations belongs only to
the association and follows almost perfectly isochrone models for PMS
stars of masses down to $\sim$ 0.3 M{\solar}. We find that this feature
has a Galactic counterpart and that the mass spectrum of the candidate PMS
stars in LH 52 seems to correspond to a Salpeter IMF with a slope $\Gamma
\simeq -1.26$ in the mass range 0.8 - 1.4 M{\solar}.
\end{abstract}


\section{Introduction}

The search for pre-main sequence (PMS) stars in young stellar systems of
the Galaxy and the Large Magellanic Cloud (LMC) has verified their
existence from both ground-based and space observations. Hillenbrand et
al. (1993)  observed the galactic open cluster NGC 6611 ``in the act of
forming intermediate-mass stars'' of 3 - 8 M{\solar}. They found that
high-mass stars of age $\sim 2$ Myr coexist with a PMS population as young
as 0.25 Myr, suggesting that the massive stars did not alter the star
formation process. Brandl et al. (1999) using VLT/ISAAC observations
revealed a PMS population in NGC 3603, which is the most massive {\sc Hii}
region in the Galaxy. Recent WFPC2 imaging with the {\em Hubble Space
Telescope} (HST) showed that the CMD position of stars fainter than
$M_{\rm V} \simeq 0$ mag indicates PMS stars with masses down to 0.1
M{\solar} with an age between 0.3 - 1 Myr (Stolte et al. 2004).

In the LMC, Brandl et al. (1996) detected with near-infrared imaging of
the central cluster of 30 Doradus (R 136) 108 ``extremely red sources'',
which are most likely PMS stars of masses down to $\sim$ 3 M{\solar}. PMS
stars in R 136 were originally revealed by Hunter et al. (1995) with
HST/WFPC2 observations. A red population with masses down to $\sim$ 1.35
M{\solar}, well tracked by PMS isochrones of ages 1 - 10 Myr, was also
discovered in R 136 by Sirianni et al. (2000) from HST/WFPC2, although no
variable extinction was taken into account (Brandl \& Andersen 2005).
HST/NICMOS observations showed that in the region of 30 Doradus there are
stars of lower mass down to 0.4 M{\solar} (Zinnecker 1998) with no
evidence for a lower mass cutoff (Brandner et al. 2001). Romaniello et al.
(2005) identified a population of objects with H{\alp} and/or Balmer
continuum excess, which they interpret as low mass ($\sim$ 1
- 2 M{\solar}) PMS stars.




As far as non-star-burst young stellar systems are concerned, (e.g.  
stellar associations), studies in our galaxy have revealed their PMS
population, but such stars {\em have not yet been detected, although they
should exist also in stellar associations of the Magellanic Clouds (MCs)}.
According to Massey et al. (1995) the regions of galactic OB associations
show evidence of PMS stars with masses between 5 - 10 M{\solar} and ages
$<$ 1 Myr. Preibisch \& Zinnecker (1999)  discovered 95 low-mass PMS stars
with an age of $\sim$ 5 Myr in the Upper Scorpius OB association and they
suggest a coevality in the formation of low- and high-mass stars. Recent
studies of the Orion OB1 association revealed its PMS population in the
mass range 0.2 - 2 M{\solar} (Sherry et al. 2004; Brice\~{n}o et al.
2005). The only comparable study in the MCs is the one around SN 1987A by 
Panagia et al. (2000), who identified several hundreds PMS stars through
their H{\alp} emission, and they found that most of them have masses in
the range 1 - 2 M{\solar} and ages between 1 - 2 and 20 Myr.

Associations in the MCs {\em are close enough for low-mass stars to be
observed} with high-resolution ground-based and space imaging, and they
{\em do not suffer from crowding or high extinction}, like the star-burst
of 30 Doradus or compact young clusters. Therefore, {\em if PMS stars are
to be found in extra-galactic stellar systems, stellar associations of the
MCs should be considered as the best places to search}. In this letter we
report the serendipitous discovery of low-mass candidate PMS stars found
with HST/WFPC2 in the Color-Magnitude Diagram (CMD) of the LMC association
LH 52 earlier published by Gouliermis, Brandner \& Henning (2005a, from
here on Paper I).

\section{PMS Stars in LMC Associations with HST}

\subsection{Observations and Data Reduction}

The WFPC2 images on the association LH 52 were taken within the HST
program GO-8134. WFPC2 data of the background LMC field in the area --
observed within the HST program GO-8059 -- are also used. Details for both
datasets are given in Paper I. Results on the Initial Mass Function (IMF)
of the general LMC field with WFPC2 observations have been recently
presented by Gouliermis, Brandner \& Henning (2005b, from here on Paper
II). These observations (HST program GO-8576), which consist of six
sequential telescope pointings of the inner LMC disk, described in Paper
II, are also used here. The photometry was performed with the package
HSTphot (Dolphin 2000), and is described in detail in Papers I and II.

\begin{figure*}[t!]
\centerline{\hbox{
\psfig{figure=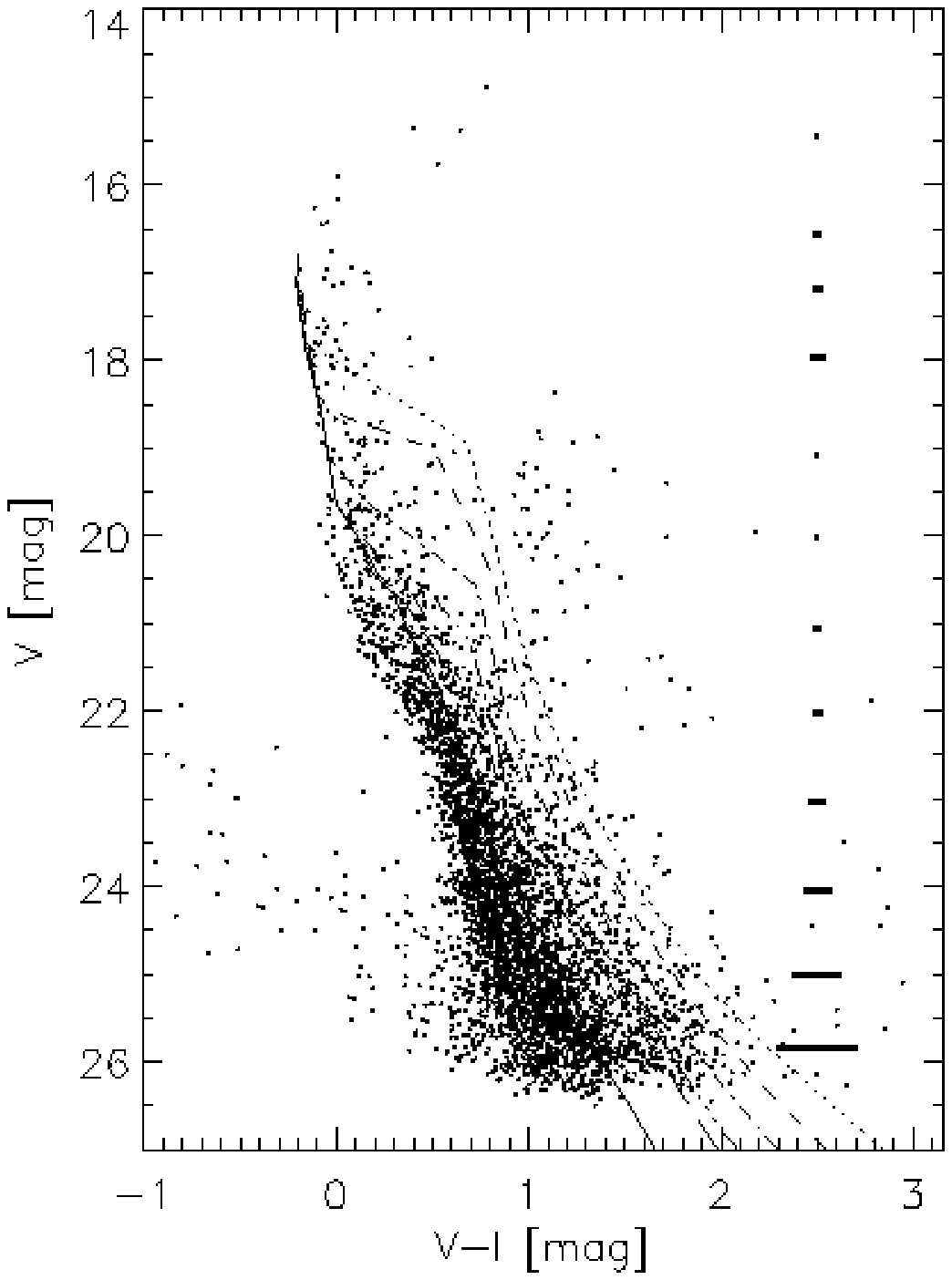,width=0.937\columnwidth,angle=0}
\psfig{figure=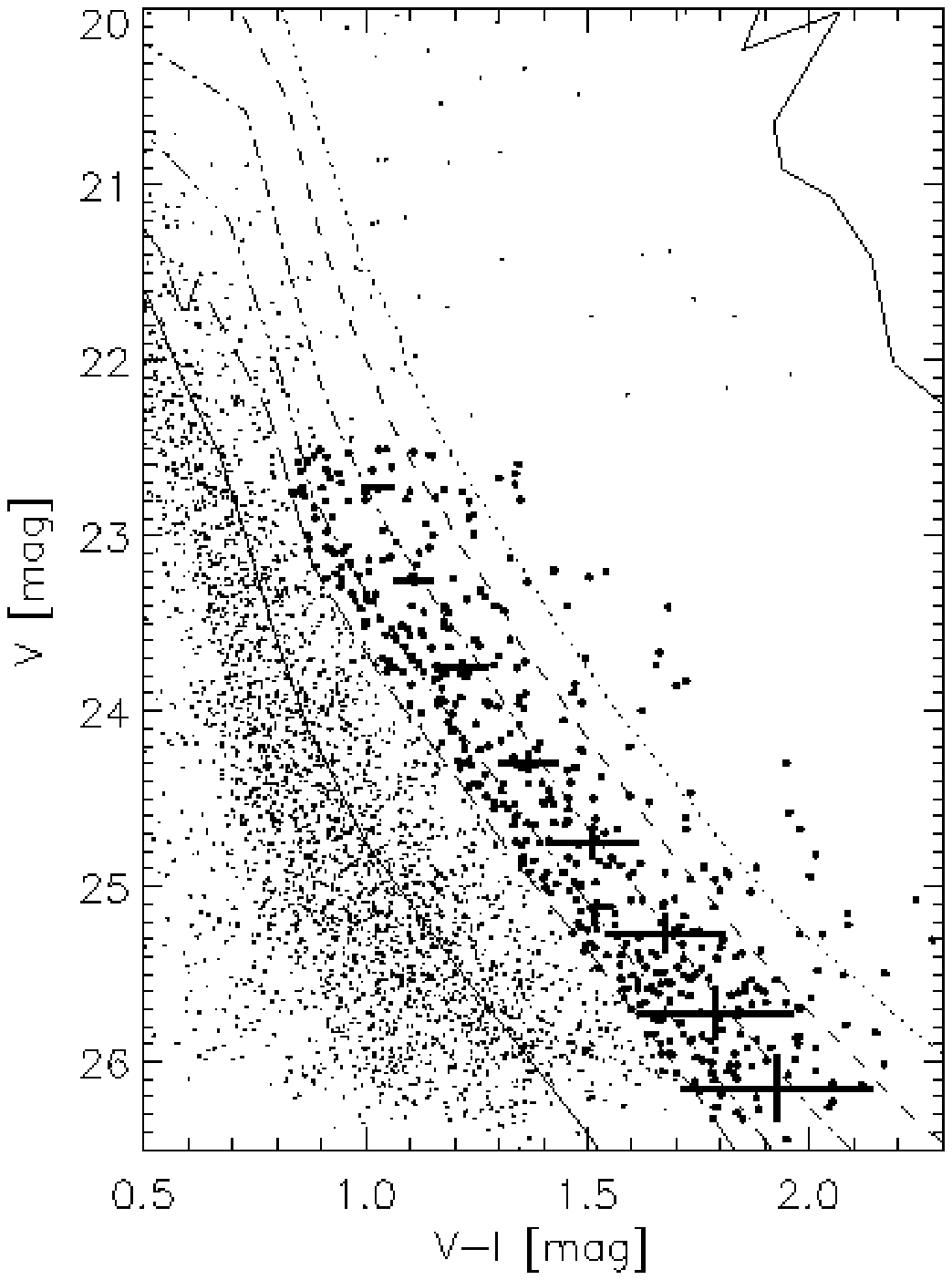,width=0.937\columnwidth,angle=0}
}}
\caption{Left: $V$, $V-I$ CMD of the stars in the observed WFPC2 field of
the association LH 52 (Paper I). Overplotted PMS isochrone models from
Siess et al. (2000) track very well the discovered faint red {\em
Sequence}, which does not seem to coincide with either a binary sequence
or main-sequence broadening due to photometric errors. Thick horizontal
lines on the right indicate typical mean color errors per magnitude.  
Right: The red fainter-most part of the CMD, where the {\em Sequence} is
located. The candidate low-mass PMS stars are represented with thick dots.
The overplotted PMS isochrones are for ages 1.5, 2.5, 5, 10 and 15 Myr and
for metallicity Z=0.004. The ZAMS and birthline are plotted with thick
lines. HSTphot fitting errors in color and magnitude are also shown.}
\label{lh52pmscmdiso}
\end{figure*}

\subsection{Discovery of candidate PMS stars}

In the $V$, $V-I$ CMD of the association LH 52 constructed in Paper I from
HST/WFPC2 observations, a secondary red sequence of faint
stars\footnote{From here on we will refer to this feature in the CMD as
the {\em Sequence}.} can be seen easily almost parallel to the lower main
sequence from the turn-off down to the detection limit (Figure 6 of Paper
I).  In Paper I we discuss the fact that this reddened {\em Sequence}
cannot be seen in the CMD of the local background field. There are two
reasons for {\em not} interpreting it as a ``normal'' CMD feature: (1) The
{\em Sequence} cannot be accounted for by binaries of main sequence stars,
because it is located more than one magnitude away from the ZAMS. (2) It
cannot be the result of broadening of the main-sequence due to photometric
errors, because in that case there would be a noisy stellar distribution
rather than a well formed sequence. The numbers of stars in the {\em
Sequence} with colors 5$\sigma$ away from the ZAMS give additional
evidence that this feature is not due to errors in the colors. For
example, the expected number of 5$\sigma$ outliers with 23.0 $\leq V \leq$
23.5 mag is $\sim$ 5, given the total number of stars in this magnitude
range, while we find that 45 stars of the {\em Sequence} are located
further than 5$\sigma$ away from the main sequence. The corresponding
numbers for stars with 23.5 $\leq V \leq$ 24.0 mag are about 9 expected
and 55 counted stars located more than 5$\sigma$ to the red of the ZAMS.
On the other hand there are facts, which support the suggestion that the
{\em Sequence} can be accounted for by a PMS population in the area of the
association LH 52.

First, the PMS evolutionary models of Siess et al. (2000) seem to trace
very well the {\em Sequence}. The WFPC2 CMD of LH 52 is shown again in
Figure \ref{lh52pmscmdiso} (left panel) with indicative PMS isochrones of
ages between 1.5 and 15 Myr overplotted. We selected $\sim$ 500 stars,
which form the {\em Sequence} and we plot with thick points their loci in
the enlarged faint red part of the CMD (Figure \ref{lh52pmscmdiso} right
panel). The limit for the brightest of the selected stars is chosen in a
way that any confusion with the stellar population of the turn-off can be
avoided. Still, PMS stars brighter than this limit are expected to exist
according to the models. The distance modulus used is $m-M = 18.5$ mag
(Panagia et al. 1991) and the isochrones are reddened for $E(B-V)  \simeq
0.075$ mag (Paper I). The excellent agreement of the {\em Sequence} with
the PMS isochrones shown in Figure \ref{lh52pmscmdiso} gives the first
evidence that these stars are of PMS nature.

\begin{figure*}[t!]
\centerline{\hbox{
\psfig{figure=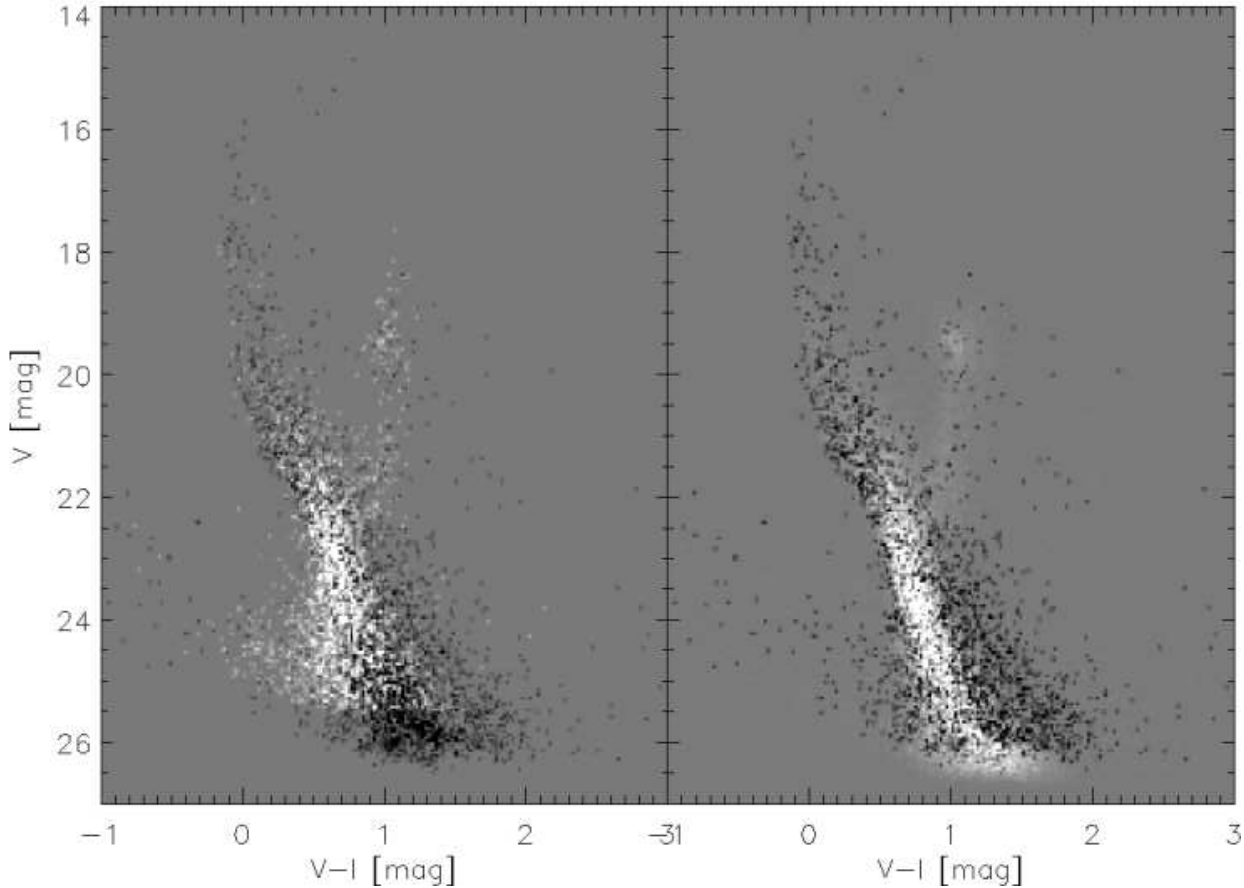,width=0.9\textwidth,angle=270}
}}
\caption{Differential Hess diagrams. Dark regions show areas of higher
density in LH 52, white regions show higher density in the local LMC
background field (left - Paper I) or in the general field at the inner
disk of LMC (right - Paper II). An excess of stars at the red faint {\em
Sequence} of candidate PMS stars in LH 52 is clearly identified.}
\label{fig_hess}
\end{figure*}

A second evidence for the PMS population of LH 52 can be seen in Figure
\ref{fig_hess}, which shows the differential Hess diagrams created by
subtracting the scaled Hess diagram of LH 52 from the ones of the close-by
general LMC field (Paper I) and of the background field of the LMC inner
disk (Paper II). The CMDs were binned in 0.02 mag intervals in $V-I$ and
0.05 mag in $V$, and normalized such that the total number of stars were
equal. The resulting images are the Hess diagrams used to produce Figure
\ref{fig_hess}. The blackest pixels represent a stellar excess in the
field of LH 52, and the whitest pixels show a stellar excess in the
background fields. The differences between the areas are dramatically
highlighted, especially in the case of the LMC disk field (right panel),
which is characterized by excellent photometry (with longer exposure
times) of a rich sample of over 80,000 stars. In the case of the close-by
field (left panel) it is shown that the lower main-sequence of this CMD,
which is broadened due to photometric errors does not cover the {\em
Sequence}. The differential Hess diagrams emphasize the fact that the
population covered by the secondary faint {\em Sequence} does not have any
counterpart in the general field of the LMC. This strengthens the
suggestion that {\em the Sequence represents PMS stars in the association
and not low-mass stars on its main-sequence}.

In addition to the discussion above, we consider the observational CMDs
recently published for the low-mass PMS stars of the Orion OB1 association
in the Galaxy. Sherry et al. (2004) presented their $BVRI$ survey of the
low-mass PMS population of the Ori OB1b sub-association. The location of
the PMS stars in their $V$, $V-I_{\rm c}$ CMD matches almost perfectly our
{\em Sequence}, if compared at the same distance. Furthermore, Brice\~{n}o
et al. (2005)  presented the CIDA variability survey of Ori OB1 and their
$V$, $V-I_{\rm c}$ CMD shows a sequence of red faint stars almost ``1
magnitude above the ZAMS''. The location in the observed CMD of their
brighter stars, which are identified as T Tauri stars with $M$ \lsim\ 2
M{\solar}, is in excellent agreement with our {\em Sequence}, if we take
the distance of the LMC into account. In both studies the comparison of
the CMD at the area of Ori OB1 with CMDs of ``control" fields away from it
proves that the PMS stars belong only to the association. This coincidence
of the loci of low-mass PMS stars in the CMD of a galactic association
with the observed faint red {\em Sequence} seen in our CMD of LH 52 gives
additional confidence that {\em this sequence accounts for the PMS
population of the association}.

\begin{figure*}[t!]
\centerline{\hbox{
\psfig{figure=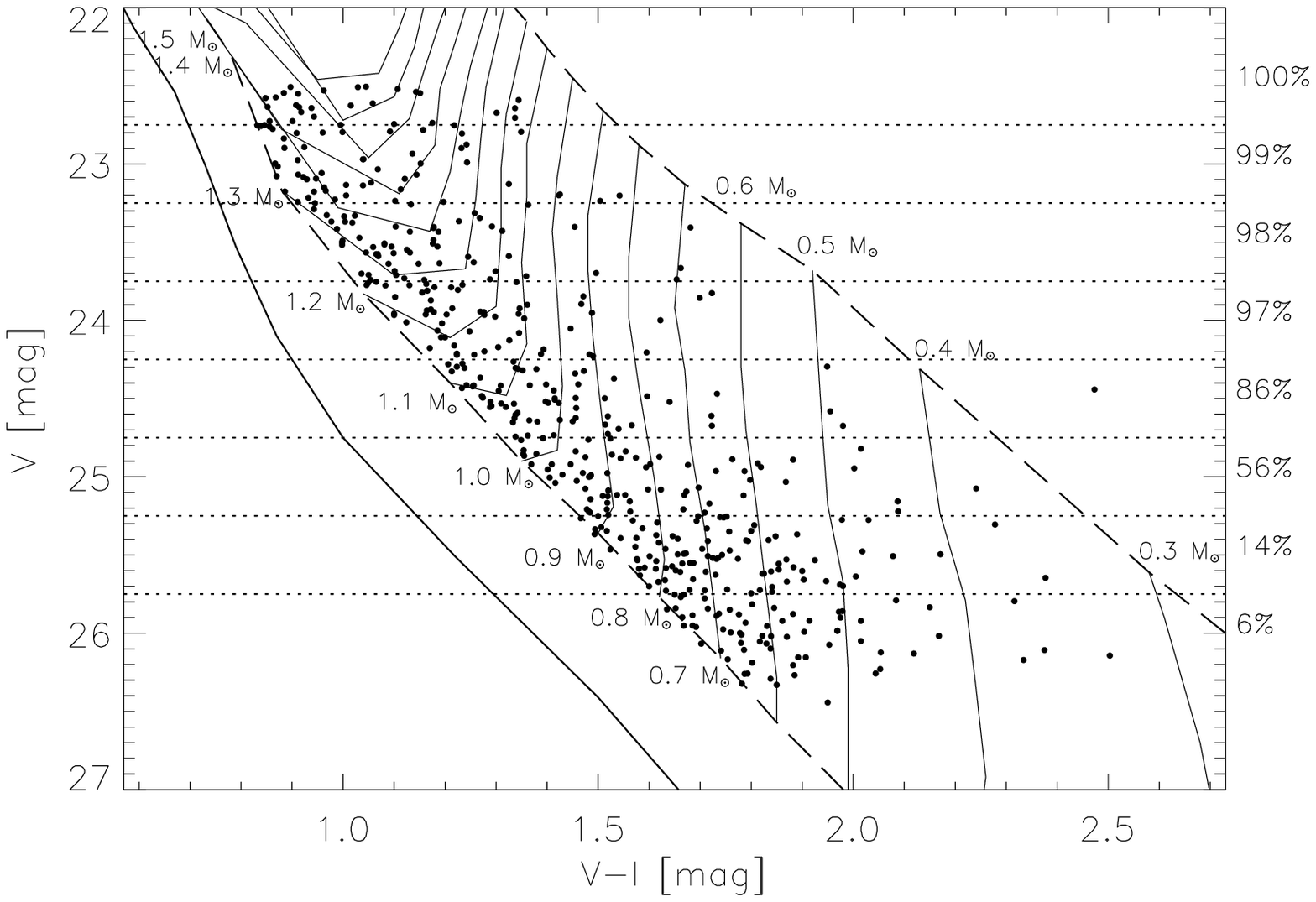,width=0.5\textwidth,angle=0}
\psfig{figure=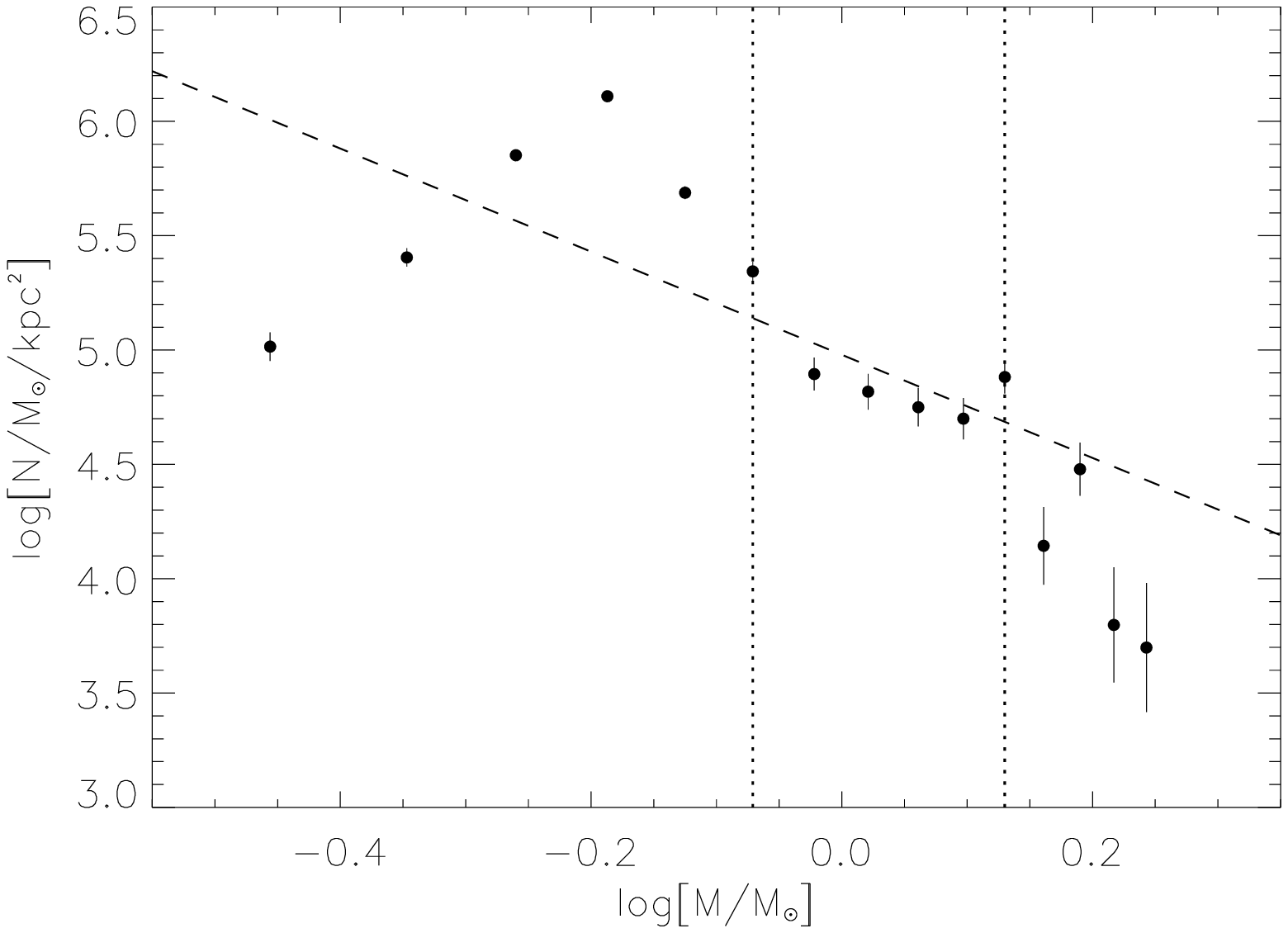,width=0.5\textwidth,angle=0}
}}
\caption{Left: PMS evolutionary tracks for masses down to $\sim$ 0.3
M{\solar} overplotted on the CMD. They are almost vertical to the $V-I$
axis, so that the use of a single M/L relation is not sufficient for the
translation of luminosities of PMS stars to masses. The corresponding mass
is indicated next to each track. $V$-magnitude completeness (in \%)  is
also given. Right: The IMS constructed by counting the candidate PMS stars
between the tracks (mass bins of 0.1 M{\solar}). The stellar numbers are
corrected for incompleteness and they are normalized to the surface of 1
kpc$^{2}$. Vertical dotted lines indicate the useful mass range of 0.8 -
1.4 M{\solar}. The corresponding IMS slope, which is computed using the 
logarithm of the number of stars normalized by the logarithm of the size 
of the mass bin, is $\gamma = -2.26 \pm 1.05$ and it is plotted with a 
dashed line.} 
\label{fig_imf}
\end{figure*}

\section{The Initial Mass Spectrum}

If we assume that the stars of the {\em Sequence} are indeed PMS stars, we
can study the low-mass PMS Initial Mass Spectrum (IMS) of LH 52. The
construction of such a distribution function would require a procedure
more thorough than the ``typical'' construction of a mass-luminosity (M/L)
relation based on isochrone models. Since PMS isochrones of different ages
are so close to each other (see Figure \ref{lh52pmscmdiso}), it is very
difficult to establish an accurate M/L relation, because it is not
possible to know the exact age of each PMS star. A more efficient process
for distributing these stars according to their masses is to count them
directly on the CMD. From the models of Siess et al. (2000) we construct
``evolutionary'' tracks (of equal mass) between isochrones of ages
starting from 0.5 Myr up to 15 Myr. These ``iso-mass" tracks, provide the
mass bins in which the PMS stars will be counted for the construction of
their IMS (Figure \ref{fig_imf} left panel).

Completeness at the magnitude ranges of our candidate PMS stars plays an
important role. As shown in Figure \ref{fig_imf} (left panel), only stars
with masses down to 0.8 M{\solar} can be used due to incompleteness. In
addition, since we strictly select the brightest limit of the {\em
Sequence} to be well below the turn-off, the numbers of counted stars in
the higher-mass bins are also incomplete.  We consider that numbers for
stars up to 1.4 M{\solar} are useful. The IMS, which is constructed by
counting stars between evolutionary tracks, is plotted in Figure
\ref{fig_imf} (right panel). Its logarithmic slope\footnote{The slope of
the IMS ($\gamma$) is related to the slope $\Gamma$ of the IMF as $\gamma
= \Gamma - 1$. Thus a Salpeter IMS has $\gamma \simeq -2.35$ (see also
Scalo 1986).}, which is $\gamma = d\log{N}/d\log{m}$, is calculated by
applying a least-square linear fit for stars within the useful mass range
of 0.8 $\leq M$/M{\solar} $\leq$ 1.4. The IMS slope is almost identical to
a Salpeter (1955) IMS, equal to $\gamma\simeq -2.26 \pm 1.05$. This slope
should be considered only as indicative due to poor statistics. Indeed,
the large uncertainty demonstrates the need for more complete data for PMS
stars in MCs associations.

\section{Final Remarks}

We provide evidence of the possible existence of PMS stars in the vicinity
of a {\em bona-fide} stellar association in the LMC. The study presented
here cannot be treated as a direct proof of their existence, but it should
be considered as an initial step for the investigation of faint red
sources located in the vicinity of star-forming MCs associations.  Recent
ACS observations of the bright association NGC 346 in the Small Magellanic
Cloud revealed ``an underlying population of 2,500 infant stars embedded
in the nebula'' (STScI press release STScI-PRC2005-04).  It is almost
certain that deep observations with HST/ACS of such systems will provide
more detailed information. Furthermore, high-resolution near-infrared
observations should be performed as follow-up to an ACS detection in the
optical and H{\alp} for the characterization of these red sources.  Such a
multi-wavelength set of data will help to characterize the PMS population
of stellar associations of the MCs, and to deal with important issues such
as {\em how long does it take for the pre-main sequence phase in MCs
associations to be completed}, and {\em how important is the role of MCs'
lower metallicities for this timescale}.


\acknowledgments

This paper is based on observations made with the NASA/ESA Hubble Space
Telescope, obtained from the data archive at the Space Telescope Science
Institute. STScI is operated by the Association of Universities for
Research in Astronomy, Inc. under NASA contract NAS 5-26555.


   
\end{document}